\documentclass[aip, amsmath,amssymb, reprint]{revtex4-1}

\usepackage{graphicx}
\usepackage{dcolumn}
\usepackage{bm}
\usepackage[utf8]{inputenc}
\usepackage[T1]{fontenc}
\usepackage{mathptmx}
\usepackage{float}

\begin{document}
\preprint{AIP/123-QED}

\title{A four-state magnetic tunnel junction switchable with spin-orbit torques}

\author{Shubhankar Das}
\email{shubha.biu.phy@gmail.com}
\affiliation{Department of Physics, Nano-magnetism Research Center, Institute of Nanotechnology and Advanced Materials, Bar-Ilan University, Ramat-Gan 52900, Israel}
\author{Ariel Zaig}
\affiliation{Department of Physics, Nano-magnetism Research Center, Institute of Nanotechnology and Advanced Materials, Bar-Ilan University, Ramat-Gan 52900, Israel}
\author{Moty Schultz}
\affiliation{Department of Physics, Nano-magnetism Research Center, Institute of Nanotechnology and Advanced Materials, Bar-Ilan University, Ramat-Gan 52900, Israel}
\author{Susana Cardoso}
\affiliation{Instituto de Engenharia de Sistemas e Computadores – Microsistemas e Nanotecnologias (INESC MN), 1000-029 Lisbon, Portugal}
\affiliation {Departamento de F\'{\i}sica, Instituto Superior Tecnico (IST), Universidade de Lisboa, 1040-001 Lisbon, Portugal}
\author{Diana C Leitao}
\affiliation{Instituto de Engenharia de Sistemas e Computadores – Microsistemas e Nanotecnologias (INESC MN), 1000-029 Lisbon, Portugal}
\affiliation {Departamento de F\'{\i}sica, Instituto Superior Tecnico (IST), Universidade de Lisboa, 1040-001 Lisbon, Portugal}
\author{Lior Klein}
\email{Lior.Klein@biu.ac.il}
\affiliation{Department of Physics, Nano-magnetism Research Center, Institute of Nanotechnology and Advanced Materials, Bar-Ilan University, Ramat-Gan 52900, Israel}

\date{\today}

\begin{abstract}
We present a magnetic tunnel junction (MTJ) where its two ferromagnetic layers are in the form of a single ellipse (SE) and two-crossing ellipses (TCE). The MTJ exhibits four distinct resistance states corresponding to the four remanent states of the TCE structure. Flowing current in an underlying Ta layer generates in the adjacent TCE structure spin-orbit torques which induce field-free switching of the four-state MTJ between all its resistance states. The demonstrated four-state MTJ is an important step towards fabricating multi-level MTJs with numerous resistance states which could be important in various spintronics applications, such as multi-level magnetic random access or neuromorphic memory.
\end{abstract}

\maketitle

Magnetic tunnel junctions (MTJs) are devices consisting of two ferromagnetic (FM) layers separated by a thin insulating layer. Commonly, the two FM layers exhibit uniaxial magnetic anisotropy and the MTJ displays two distinguishable states, with the FM layers being oriented either parallel or antiparallel. The two discrete magnetic states of the MTJ correspond to two resistance states which are the basis for a two-state memory bit used, for instance, in magnetic random access memory (MRAM)\cite{Bhatti}.

While two-state memory bits are ubiquitous, there are increasing efforts to fabricate multi-level memory bits which can store more than two states in different memory technologies including FLASH \cite{Shibata,Coughlin}, phase change \cite{Suri,Athmanathan} and resistive memory \cite{Xu,Prezioso}. These efforts are driven by the demand for denser memory needed to cope with the accelerated rate at which we generate data and the realization that increasing the density by decreasing the memory bit size is approaching intrinsic barriers such as superparamagnetism in the case of MRAM \cite{Bean,Cowburn,Skumryev} or tunneling of charge carriers in the case of FLASH.

Multi-level memory bits are attractive not only for increasing memory density but also because they can be used for various types of memory including memristors \cite{Strukov,Wang,Yang,Fukami_natmat,Olejnik} and for being used for neuromorphic computing  \cite{Jo,Lequeux,Kurenkov_advmat,Grollier,Kurenkov}. Notably, memory cells having merely  14-20  states were shown to be sufficient for pattern classification tasks, such as recognition of handwritten digits \cite{Querlioz}.

Irrespective of the potential application of a multi-level MTJ, it is important that the method used for switching between the MTJ states is scalable and non-detrimental. These two requirements are met when the switching is achieved by spin-orbit torques (SOTs) which are generated by spin currents injected into a FM layer due to conversion of charge current into spin current in an adjacent heavy metal layer \cite{Miron,Liu,Pai,Yu,Garello,Hao,Hung,Oh,Aradhya,Fukami,Xu_SCIREP,Chen,Baek,Das_swit,Das_domino,Das_exponential}.

Here we present the realization of a four-state MTJ cell fabricated by using FM layers in the form of a single ellipse (SE) and two crossing ellipses (TCE) as top and bottom layers, respectively. The top layer exhibits uniaxial magnetic anisotropy and it acts as a pinned layer with unchanging magnetization orientation, whereas the bottom layer exhibits effective bi-axial magnetic anisotropy in the crossing area of the ellipse and it acts as a free layer which can switch between four remanent states. Consequently, the MTJ has four remanent magnetic configurations which give rise to four resistance states.  The MTJ is on top of a Ta cross which allows its switching by SOTs induced by flowing current in the underlaying Ta layer. As the method presented here for increasing the number of resistance states of a MTJ can be extrapolated for achieving a much larger number of resistance states, the demonstrated four-states MTJ which is switchable with SOTs is an important step towards realization of multi-level MRAM. In addition, multi-level MTJs may be important in other applications such as neural computation.

Our devices are fabricated using heterostructures of $\beta$-Ta(10)/Ni$_{0.8}$Fe$_{0.2}$(2)/Al$_2$O$_3$(1.2)/Ni$_{0.8}$Fe$_{0.2}$(4)/Ta(5), where the numbers in parenthesis are in nanometer. The layers are deposited on thermally oxidized Si substrate using a N3000 ion-beam sputtering deposition system (the details of the deposition is described in Ref. \cite{Cardoso,Knudde_IBS}). In particular, the AlO$_X$ tunnel barriers were prepared by a two-step deposition/oxidation process. First, Al(0.7 nm) is deposited and oxidized by remote plasma for 15 sec (mixture of 20 sccm oxygen and 4 sccm argon), followed by a repetition of  Al(0.5 nm) deposition and 15 sec oxidation \cite{Knudde}.

\begin{figure*}[!]
\begin{center}
\includegraphics [trim=0cm 0cm 0cm 0cm, angle = 0, width=17cm,angle=0]{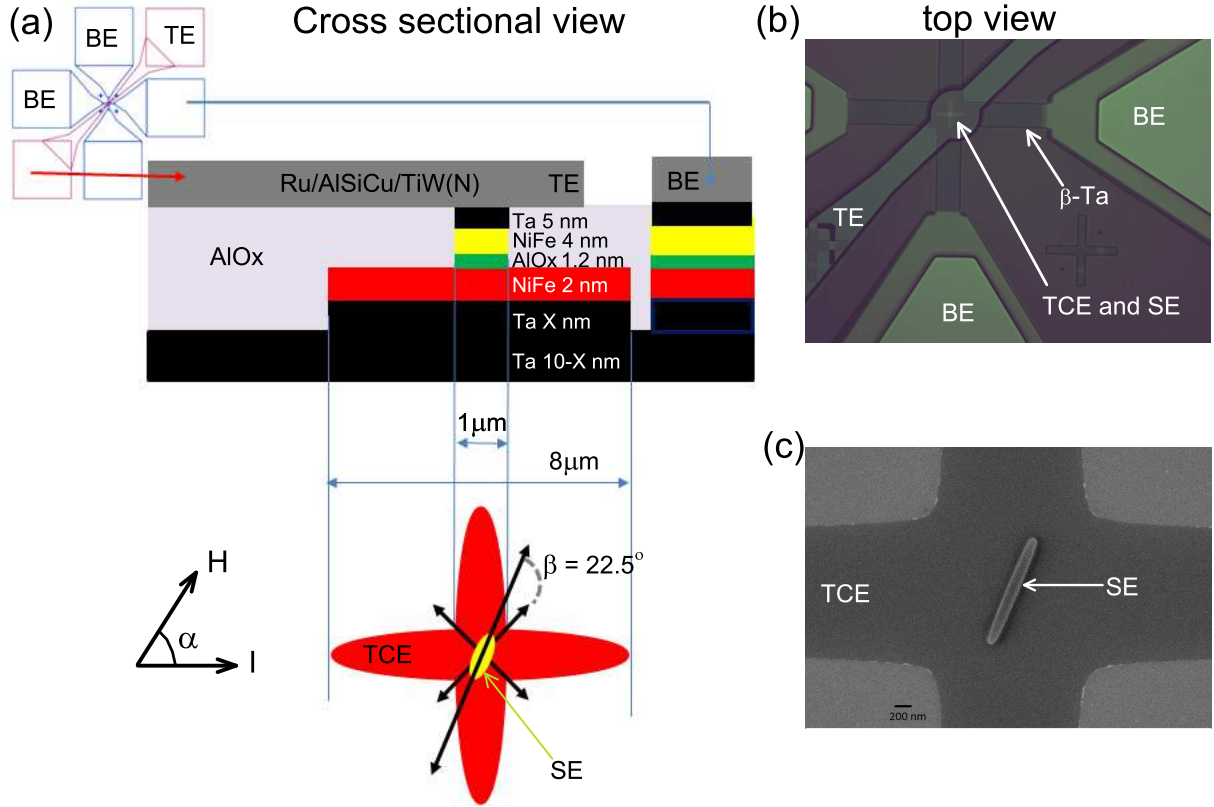}
\end{center}
\caption {\label{Schematics_MTJ} (a) Schematic of the MTJ, including (top left panel) the mask layout with the bottom electrode (BE) of $\beta$-Ta patterned into a crossed shape (the numbering denotes the contact pads) and the top electrode (TE), (center panel) the cross sectional view of the full MTJ stack and (bottom panel) the relative orientations of the TCE and the SE defined at 22.5$^{\circ}$. The easy axes of TCE are denoted as M$_1$, M$_2$, M$_3$ and M$_4$. (b) Top-view optical image of the overall device, showing in the middle the TE over the TCE and SE. (c) SEM image of the MTJ area where the top SE is fabricated at the overlap area of the bottom TCE.}
\end{figure*}

The bottom $\beta$-Ta layer is patterned in the form of a cross, with a crossing area of 6$\times$6 $\mu$m$^2$ by laser lithography followed by Ar-ion milling. At the crossing area, the bottom NiFe layer is patterned by electron beam lithography and Ar-ion milling in the form of TCE, with the principal axes of 1 and 8 $\mu$m long.
A nominal resist thickness of 500 nm was used, with optimized energy and development time, to minimize backscattering effects at the crossing area, while ensuring well defined ellipses tips \cite{Leitao}. The bottom $\beta$-Ta layer is then passivated first with Al$_2$O$_3$(9 nm) using RF sputtering. Lift-off is performed to access the TCE. Then, a second e-beam lithography followed by a two-angle Ar-ion milling \cite{Leitao}, is used to define the Al$_2$O$_3$ junction and top NiFe layers into a SE of dimension 150 nm$\times$900 nm in the overlap area of the TCE. The long axis of the SE makes an angle of 22.5$^{\circ}$ with one of the principal axes of the TCE (Fig. \ref{Schematics_MTJ}). The final passivation is performed with Al$_2$O$_3$(50 nm) and lift-off to open top via to the nanopillar. Top electrode is defined by laser lithography and lift off of ion-beam deposited Ru(40 nm) and sputtered Al$_{98.5}$Si$_{1}$Cu$_{0.5}$(300 nm)/Ti$_{10}$W$_{90}$(\rm N$_2$) (N$_2$ denotes that during the deposition of the layer the atmosphere was mostly of Nitrogen that incorporates N$_2$ in the film to render it inert and work as a better protective layer). The schematics and the scanning electron microscopy (SEM) images are shown in Fig. \ref{Schematics_MTJ}. All the measurements are done at room temperature using a home-made system consisting of a Helmholtz coil and a sample rotator with resolution of 0.03º.

As shown in previous works, a SE structure exhibits uniaxial magnetic anisotropy with the easy axis along the long axis of the ellipse \cite{Mor,Das_swit}, whereas TCE exhibit bi-axial magnetic anisotropy with the easy axes in between the long axes of the ellipses \cite{Telepinsky_JAP,Das_domino}. As in our experiments we want the magnetization of the SE to remain fixed, it is designed so that its minimum switching field ($>$ 100 Oe)\cite{Mor} would be much larger than the minimum switching field of the TCE ($\sim$ 5 Oe)

\begin{figure}[h]
\begin{center}
\includegraphics [trim=0cm 0cm 0cm 0cm, angle = 0, width=8.5cm,angle=0]{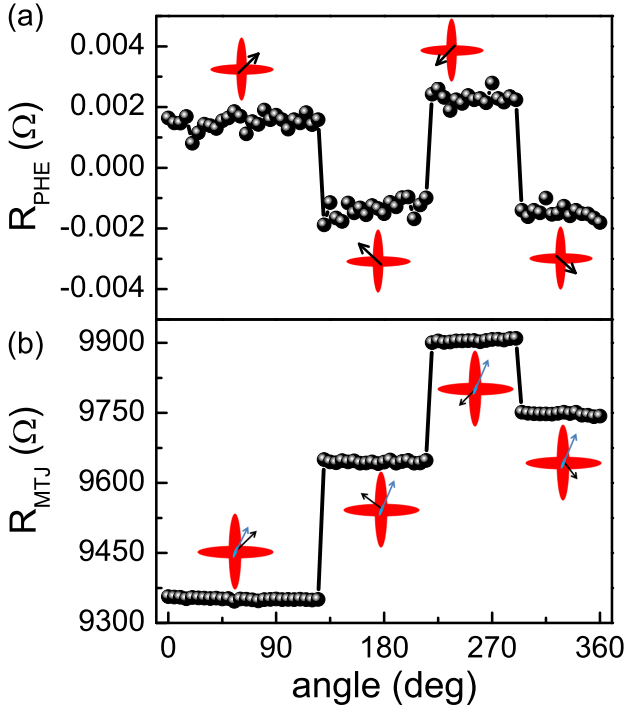}
\end{center}
\caption {\label{Magnetic_characterization} (a) and (b) R$_\text{PHE}$ and R$_\text{MTJ}$, respectively, as a function of the angle $\alpha$ at which a field of 7 Oe is applied and removed. The presented data are taken at zero field for each field direction. The illustrations show the direction of magnetization of top SE structure (blue arrow), the bottom TCE structure (black arrow).}
\end{figure}

Fig. \ref{Magnetic_characterization} presents magnetic characterization of our device performed by measuring planar Hall resistance (R$_\text{PHE}$) and the MTJ resistance (R$_\text{MTJ}$). R$_\text{PHE}$ is measured by driving a current of 100 $\mu$A along one of the arms of the Ta cross and measuring the voltage across the other Ta arm. R$_\text{MTJ}$ is measured with a current of 5 $\mu$A through the MTJ (see Fig 1). Fig. \ref{Magnetic_characterization}(a) and (b) show  R$_\text{PHE}$ and R$_\text{MTJ}$ measured after a field of 7 Oe is applied and removed at different angles $\alpha$ between current and field direction. The chosen field is sufficient to switch the TCE between its four remanent states but too small to flip the remanent magnetization of the SE. We note four different values for  R$_\text{MTJ}$ and two different values for R$_\text{PHE}$.   R$_\text{MTJ}$ depends on the relative orientation of the magnetization of the FM layers and is given by \cite{Telepinsky_APL};
\begin{equation}
R_\text{MTJ}(\beta) = R_\text{AV} - \frac{1}{2} \Delta R\: cos\:\beta
\label{MTJ_resistance}
\end{equation}
where $\beta$ is the angle between the magnetization direction of the two layers, R$_\text{AV}$ = (R$_\text{P}$ + R$_\text{AP}$)/2, R$_\text{P}$ and R$_\text{AP}$ are the resistance of the MTJ when the magnetization of the two layers is in parallel and anti-parallel states, respectively, and $\Delta$R = R$_\text{AP}$ - R$_\text{P}$. The four states of R$_\text{MTJ}$ follow the above mentioned Eq. \ref{MTJ_resistance}. The exact switching positions between the remanent states in both Fig. \ref{Magnetic_characterization}(a) and (b) strongly indicate that the four R$_\text{MTJ}$ values correspond to the four states of the TCE. R$_\text{PHE}$ shows two resistance states due to the symmetry of R$_\text{PHE}$ under magnetization inversion. To obtain maximum separation between the states, the SE long axis is rotated 22.5$^0$ from the principal axis of the TCE (see Fig. \ref{Schematics_MTJ}(c)). The lack of precise equidistance between R$_\text{MTJ}$ values can be explained by slight misalignment during fabrication.

\begin{figure}[!]
\centering
\includegraphics [trim=0cm 0cm 0cm 0cm, angle = 0, width=8.5cm,angle=0]{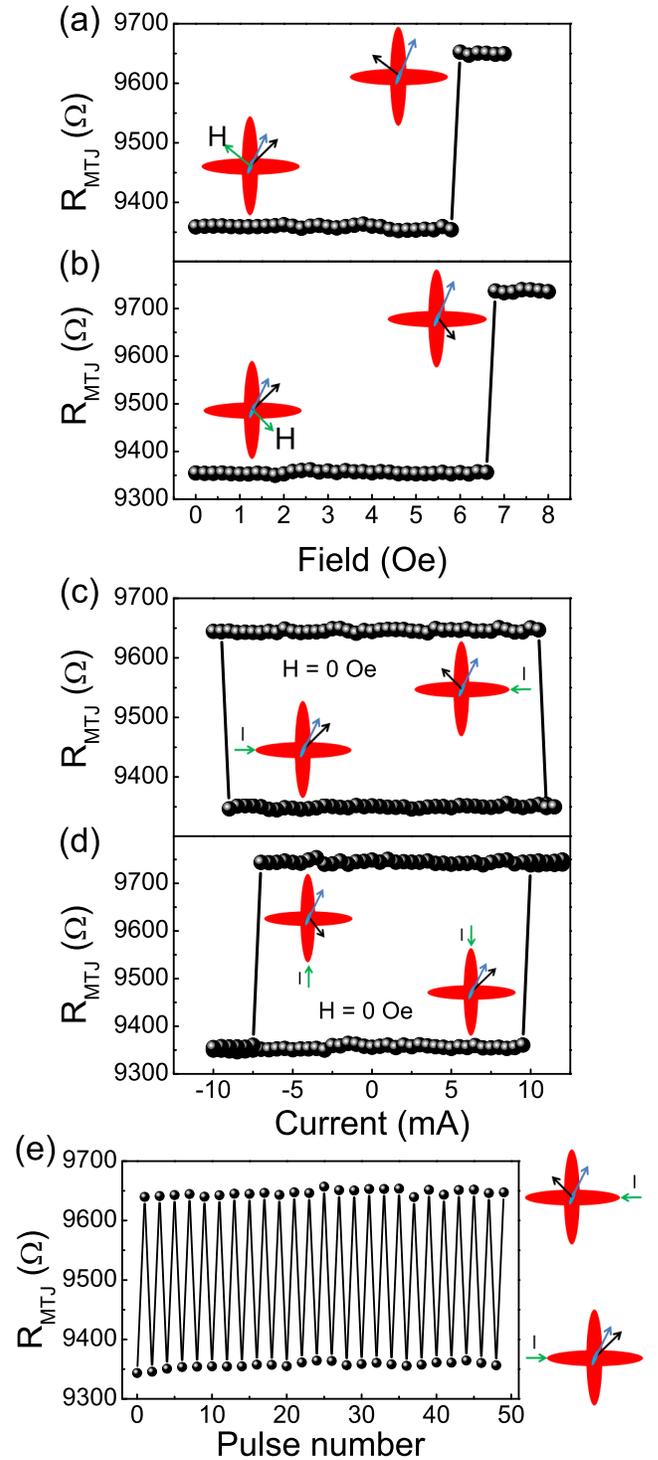}
\caption {\label{SOT_switching}(a) and (b) show field-induced switching between different remanent states. (c) and (d) Field-free SOT-induced switching between two remanent states of MTJ by flowing current in horizontal and vertical Ta-strips, respectively. (e) Successive switchings between two states, shown in (a), by flowing 50 pulses of amplitude 12 mA in opposite directions in the horizontal Ta-strip. The illustrations show the direction of magnetization of top SE structure (blue arrow) and the bottom TCE structure (black arrow) and the direction of current.}
\end{figure}

We now turn to explore switchings between the four resistance states  of the MTJ. Fig. \ref{SOT_switching}(a) and (b) show switchings between different remanent states by external applied fields. For every field value, the measurement is performed  after switching off the field. Fig. \ref{SOT_switching}(c) shows reversible switchings between two states by flowing current in the underlying horizontal Ta-strip in the absence of any external field. The current in the Ta-strip affects only the TCE structure, keeping the magnetization of the SE unaffected. The measurements are performed by flowing pulses of various amplitudes through the Ta-strip and after each pulse we measure  R$_\text{MTJ}$ with a current of 5 $\mu$A, which does not affect the magnetization in both structures. Flowing current in the vertical strip, yields reversible switchings between the two other remanent states (Fig. \ref{SOT_switching}(d)). Considering current shunting between the Ta-strip and the TCE, switching current value (I$_\text{SW}$) of 10 mA corresponds to current density of $\sim$ 1.5$\times$10$^7$ A/cm$^2$ in the Ta layer. The reproducibility of the switchings between the states is demonstrated by flowing 50 pulses of 12 mA in alternating directions in the horizontal Ta-strip (see Fig. \ref{SOT_switching}(e)). Table I shows the response of all initial states to all currents in a given Ta-strips (blank boxes mean no response).

\begin{table}[]
\begin{center}\caption{The achieved final states for all initial states (left column) after applying all possible current directions (top row). The denotions of the states and the current pads are shown in Fig. \ref{Schematics_MTJ}.} \label{TABLE I}
\begin{tabular} {|c|c|c|c|c|c|c|}
\hline
Current direction$\rightarrow$ &  I (2 to 4) & I (4 to 2) & I (1 to 3) & I (3 to 1)  \\
Initial states &   &  &  &       \\
$\downarrow$ &   &  &  &       \\
\hline
M$_1$ &  M$_4$ &  & M$_2$ &    \\
\hline
M$_2$ & M$_3$  &  &  &  M$_1$ \\
\hline
M$_3$ &   & M$_2$ &  & M$_4$  \\
\hline
M$_4$ &   & M$_1$ & M$_3$ &    \\
\hline

\end{tabular}
\end{center}
\end{table}

We note that although the switching depends both on the initial state and the current direction, all states can be achieved by applying two subsequent current pulses. State M$_1$ is achieved by applying I(4 to 2) followed by I(3 to 1); state M$_2$ by I(4 to 2) followed by I(1 to 3); state M$_3$ by I(2 to 4) followed by I(1 to 3) and state M$_4$ by I(2 to 4) followed by I(3 to 1).

The switching of the MTJ is actually the switching of the TCE structure due to current flowing in the underlying Ta layer. The switching of TCE permalloy structures on top a Ta layer (not part of a MTJ) has been previously studied.\cite{Das_swit,Das_domino} Using harmonic Hall measurements and micro-magnetic simulations, it was shown that spin-orbit torques  induced by the current flowing in the Ta layer play a dominant role in the switching. Here also in all the switching geometries, the direction of the Oersted field generated by the flowing current in the Ta-strip cannot be solely responsible for the switching, which confirms the SOTs as being vital for the switchings. We note that for current driven in the horizontal Ta-strip, switching occurs when the angle of the initial magnetization in the overlap area and the direction of current ($\gamma$) is 45 deg, whereas for current driven in the vertical Ta-strip, switching occurs when $\gamma$ is 135 deg. Possible reasons for the difference in the behavior can be due to both the N\'{e}el "orange-peel" coupling between the ferromagnetic layers of the MTJ\cite{Schrag,Tegen} and stray magnetic field created by the top SE at the overlap area of TCE, which are asymmetric relative to the two Ta strips. We note that while "orange-peel" coupling is ferromagnetic, the stray field in our configuration favors antiparallel alignment. Whereas the average stray magnetic field at the overlap area of TCE can be estimated as ~2.9 Oe by using an analytical solution\cite{Herbert}, more study is required to quantify the "orange peel" coupling strength. We note that the effect of such couplings is probably manifested also in Fig. \ref{SOT_switching}(d) where the threshold current needed to switch to a more anti-parallel alignment is higher than the reversed switching. While the presence of "orange-peel" coupling and stray field do not obstruct the realization of four states in our device, we note that the effect of both of them can be reduced and even diminished.  The "orange-peel" coupling can be reduced by increasing the thickness of the insulating barrier and/or decreasing the roughness of the interfaces, whereas the magnitude of the stray field can be decreased with a synthetic antiferromagnetic pinned layer structure on the top of the SE. We note that in devices with a single ferromagnetic layer where such couplings are absent (see Ref. 35), the switching is symmetric with respect to current flow in either horizontal or vertical Ta-strip.

The method presented here for achieving four resistance states per MTJ can be extrapolated to achieve a much larger number of resistance states. As shown recently, a structure consisting of N crossing ellipses may support up to 2$^\text{2N}$ states\cite{Das_exponential}; therefore, using in a MTJ relatively simple structures of up to N=5 would increase the number of resistance states quite considerably. We note that structures of up to N=5 were previously studied and that simulations indicate 70 resistance states of a MTJ consisting of layers of N=4 and N=5 crossing ellipses structures while considering only a small subset of accessible remanent states\cite{Das_domino}. To distinguish between the different resistance states, particularly when their number is significantly increased, it would be important to use MTJs with higher magnetoresistance. This may be achieved by using a different insulating layer (e.g. MgO) and/or different FM layers (e.g. CoFeB).

In summary, we present a MTJ that supports four remanent states which correspond to four discrete resistance values. Furthermore, we demonstrate that the MTJ can be switched repeatedly between its four states using field-free spin-orbit torques induced by flowing current in an underlying Ta layer. An efficiently switchable multi-level MTJ is attractive for high-density multi-level MRAM and other intriguing applications such as magnetic memristors\cite{Wang,Lequeux} and magnetic neuromorphic computing\cite{Lequeux,Querlioz}.

L. K. acknowledges support by the Israel Science Foundation founded by the Israel Academy of Sciences and Humanities (533/15). DC Leitao acknowledges financial support through FSE/POPH. This work was partially supported by the project LISBOA-01-0145-FEDER-031200, PTDC/NAN-MAT/31688/2017, the National Infrastructure Roadmap MicroNanoFab@PT - NORTE-01-0145-FEDER-22090 and FCT funding of the Research Unit INESC MN (UID/05367/2020) through plurianual BASE and PROGRAMATICO financing.

The data that support the findings of this study are available from the corresponding author upon reasonable request.

\bigskip
\noindent
\large{\textbf{References}}


\begin{thebibliography}{100}

\bibitem{Bhatti}
S. Bhatti, R. Sbiaa, A. Hirohata, H. Ohno, S. Fukami, and S. N. Piramanayagam, Mater. Today \textbf{20,} 530–548 (2017).

\bibitem{Shibata}
N. Shibata, H. Maejima, K. Isobe, K. Iwasa, M. Nakagawa, M. Fujiu, T. Shimizu, M. Honma, S. Hoshi, T. Kawaai et al., IEEE Journal of Solid-State Circuits \textbf{43,} 929-937, (2008).

\bibitem{Coughlin}
T. Coughlin, IEEE Consumer Electronics Magazine \textbf{6,} 126-133 (2017).

\bibitem{Suri}
M. Suri, O. Bichler, D. Querlioz, B. Traore, O. Cueto, L. Perniola, V. Sousa, D. Vuillaume, C. Gamrat, and B. DeSalvo, J. App. Phys. \textbf{112,} 054904 (2012).

\bibitem{Athmanathan}
A. Athmanathan, M. Stanisavljevic, N. Papandreou, H. Pozidis, and E. Eleftheriou, IEEE Journal on Emerging and Selected Topics in Circuits and Systems \textbf{6,} 87-100 (2016).

\bibitem{Xu}
C. Xu, D. Niu, N. Muralimanohar, N. P. Jouppi, and Y. Xie, Design Automation Conference (DAC), 2013 50th ACM/EDAC/IEEE, pp. 1-6, IEEE, 2013.

\bibitem{Prezioso}
M. Prezioso, F. Merrikh-Bayat, B. D. Hoskins, G. C. Adam, K. K. Likharev, and D. B. Strukov, Nature, \textbf{521,} 61–64 (2015).

\bibitem{Bean}
C. P. Bean, and J. D. Livingston, J. Appl. Phys. \textbf{30,} S120-S129 (1959).

\bibitem{Cowburn}
R. P. Cowburn, J. Appl. Phys. 93, 9310 (2003).

\bibitem{Skumryev}
V. Skumryev, S. Stoyanov, Y. Zhang, G. Hadjipanayis, D. Givord, and J. Nogues, Nature \textbf{423,} 850-853 (2003).

\bibitem{Strukov}
D. B. Strukov, G. S. Snider, D. R. Stewart, and R. S. Williams, Nature, \textbf{453,} 80–83 (2008).

\bibitem{Wang}
X. Wang, Y. Chen, H. Xi, H. Li, and D. Dimitrov, IEEE electron device letters \textbf{30,} 294-297 (2009).

\bibitem{Yang}
J. J. Yang, D. B. Strukov, and D. R. Stewart, Nat. Nanotechnol. \textbf{8,} 13–24 (2013).

\bibitem{Fukami_natmat}
S. Fukami, C. Zhang, S. DuttaGupta, A. Kurenkov, and H. Ohno, Nat. Mater. 15, 535 (2016).

\bibitem{Olejnik}
K. Olejn\'{\i}k, V. Schuler, X. Marti, V. Novak, Z. Kaspar, P. Wadley, R. P. Campion, K. W. Edmonds, B. L. Gallagher, J. Garces, et al., Nat. Commun. \textbf{8,} 15434 (2017).

\bibitem{Jo}
S. H. Jo, T. Chang, I. Ebong, B. B. Bhadviya, P. Mazumder, and W. Lu, Nano Letter \textbf{10,} 1297–1301 (2010).

\bibitem{Lequeux}
S. Lequeux, J. Sampaio, V. Cros, K. Yakushiji, A. Fukushima, R. Matsumoto, H. Kubota, S. Yuasa, and J. Grollier, Scientific Reports \textbf{6,} 31510 (2016).

\bibitem{Kurenkov_advmat}
A. Kurenkov, S. DuttaGupta, C. Zhang, S. Fukami, Y. Horio, and Hideo Ohno, Adv. Mater. \textbf{31,} 1900636 (2019).

\bibitem{Grollier}
J. Grollier, D. Querlioz, K. Y. Camsari, K. Everschor-Sitte, S. Fukami, and M. D. Stiles, Nat. Electron. \textbf{3,} 360-370 (2020).

\bibitem{Kurenkov}
A. Kurenkov, S. Fukami, and H. Ohno, J. Appl. Phys. \textbf{128,} 010902 (2020).

\bibitem{Querlioz}
D. Querlioz, O. Bichler, A. F. Vincent, and C. Gamrat, Proceedings of the IEEE \textbf{103,} 1398–1416 (2015).

\bibitem{Miron}
I. M. Miron, K. Garello, G. Gaudin, P.-J. Zermatten, M. V. Costache, S. Auffret, S. Bandiera, B. Rodmacq, A. Schuhl, and P. Gambardella, Nature \textbf{476,} 189–193 (2011).

\bibitem{Liu}
L. Liu, C.-F. Pai, Y. Li, H. W. Tseng, D. C. Ralph, and R. A. Buhrman, Science \textbf{336,} 555-558 (2012).

\bibitem{Pai}
C.-F. Pai, L. Liu, Y. Li, H. W. Tseng, D. C. Ralph, and R. A. Buhrman, Appl. Phys. Lett. \textbf{101,} 122404 (2012).

\bibitem{Yu}
G. Yu, P. Upadhyaya, Y. Fan, J. G. Alzate, W. Jiang, K. L. Wong, S. Takei, S. A. Bender, L.-T. Chang, Y. Jiang et al., Nat. Nanotechnol. \textbf{9,} 548-554 (2014).

\bibitem{Garello}
K. Garello, C. O. Avci, I. M. Miron, M. Baumgartner, A. Ghosh, S. Auffret, O. Boulle, G. Gaudin, and P. Gambardella, Appl. Phys. Lett. \textbf{105,} 212402 (2014).

\bibitem{Hao}
Q. Hao and G. Xiao, Phys. Rev. B \textbf{91,} 224413 (2015).

\bibitem{Hung}
Y.-M. Hung, L. Rehm, G. Wolf, and A. D. Kent, IEEE Mag. Lett. \textbf{6,} 3000504 (2015).

\bibitem{Oh}
Y.-W. Oh, S-h. C. Baek, Y. M. Kim, H. Y. Lee, K.-D. Lee, C.-G. Yang, E.-S. Park, K.-S. Lee, K.-W. Kim, G. Go et al., Nat. Nanotechnol. \textbf{11,} 878–885 (2016).

\bibitem{Aradhya}
S. V. Aradhya, G. E. Rowlands, J. Oh, D. C. Ralph, and R. A. Buhrman, Nano Lett. \textbf{16,} 5987–5992 (2016).

\bibitem{Fukami}
S. Fukami, T. Anekawa, C. Zhang, and H. Ohno, Nat. Nanotechnol. \textbf{ 11,} 621-625 (2016).

\bibitem{Xu_SCIREP}
Y. Xu, Y. Yang, K. Yao, B. Xu, and Y. Wu, Sci. Rep. \textbf{6,} 26180 (2016).

\bibitem{Chen}
W. Chen, L. Qian, and G. Xiao, Sci. Rep. \textbf{8,} 8144 (2018).

\bibitem{Baek}
S. C. Baek, V. P. Amin, Y.-W. Oh, G. Go, S.-J. Lee, G.-H. Lee, K.-J. Kim, M. D. Stiles, B.-G. Park, and K.-J. Lee, Nat. Mater. \textbf{17,} 509–513 (2018).

\bibitem{Das_swit}
S. Das, L. Avraham, Y. Telepinsky, V. Mor, M. Schultz, and L. Klein, Sci. Rep. \textbf{8,} 15160 (2018).

\bibitem{Das_domino}
S. Das, A. Zaig, H. Nhalil, L. Avraham, M. Schultz, and L. Klein, Sci. Rep. \textbf{9,} 20368 (2019).

\bibitem{Das_exponential}
S. Das, A. Zaig, M. Schultz, and L. Klein, Appl. Phys. Lett. \textbf{116,} 262405 (2020).

\bibitem{Cardoso}
S. Cardoso, V. Gehanno, R. Ferreira and P. P. Freitas, IEEE Trans. Magn., \textbf{35,} 2952-2954 (1999).

\bibitem{Knudde_IBS}
S. Knudde, D. C. Leitao, S. Cardoso, and P. P. Freitas, J. Phys. D: Appl. Phys. \textbf{50,} 165001 (2017).

\bibitem{Knudde}
S. Knudde, G. Farinha, D. C. Leitao, R. Ferreira, S. Cardoso and P.P. Freitas, J. Magn. Magn. Mater. \textbf{412,} 181-184 (2016).

\bibitem{Leitao}
D. C. Leitao, E. Paz, A. V. Silva, A. Moskaltsova, S. Knudde, F. L. Deepak, R. Ferreira, S. Cardoso, and P. P. Freitas, IEEE Trans. on Magn. \textbf{50,} 4410508 (2014).

\bibitem{Mor}
V. Mor, M. Schultz, O. Sinwani, A. Grosz, E. Paperno, and L. Klein, Appl. Phys. Lett. \textbf{111,} 07E519 (2012).

\bibitem{Telepinsky_JAP}
Y. Telepinsky, V. Mor,  M. Schultz, and L. Klein, J. App. Phys. \textbf{111,} 07C715 (2012).

\bibitem{Telepinsky_APL}
Y. Telepinsky, V. Mor, M. Schultz, Y.-M. Hung, A. D. Kent, and L. Klein, App. Phys. Lett. \textbf{108,} 182401 (2016).

\bibitem{Schrag}
B. D. Schrag, A. Anguelouch, S. Ingvarsson, G. Xiao, Y. Lu, P. L. Trouilloud, A. Gupta, R. A. Wanner, W. J. Gallagher, P. M. Rice, and S. S. P. Parkin, Appl. Phys. Lett. \textbf{77,} 2373 (2000).

\bibitem{Tegen}
S. Tegen, I. M\"{o}nch, J. Schumann, H. Vinzelberg, and C. M. Schneider, J. Appl. Phys. \textbf{89,} 8169 (2001).

\bibitem{Herbert}
R. Engel-Herbert and T. Hesjedal, J. Appl. Phys. \textbf{97,} 074504 (2005).



\end{thebibliography}

\end{document}